# *PhotoGAN*: Generative Adversarial Neural Network Acceleration with Silicon Photonics

Tharini Suresh, Salma Afifi, and Sudeep Pasricha
Department of Electrical and Computer Engineering
Colorado State University, Fort Collins, CO
{tharini, salma.afifi, sudeep}@colostate.edu

*Abstract*— Generative Adversarial Networks (GANs) are at the forefront of AI innovation, driving advancements in areas such as image synthesis, medical imaging, and data augmentation. However, the unique computational operations within GANs, such as transposed convolutions and instance normalization, introduce significant inefficiencies when executed on traditional electronic accelerators, resulting in high energy consumption and suboptimal performance. To address these challenges, we introduce *PhotoGAN*, the first silicon-photonic accelerator designed to handle the specialized operations of GAN models. By leveraging the inherent high throughput and energy efficiency of silicon photonics, *PhotoGAN* offers an innovative, reconfigurable architecture capable of accelerating transposed convolutions and other GAN-specific layers. The accelerator also incorporates a sparse computation optimization technique to reduce redundant operations, improving computational efficiency. Our experimental results demonstrate that *PhotoGAN* achieves at least 4.4× higher GOPS and 2.18× lower energy-per-bit (EPB) compared to state-of-the-art accelerators, including GPUs and TPUs. These findings showcase *PhotoGAN* as a promising solution for the next generation of GAN acceleration, providing substantial gains in both performance and energy efficiency.

## I. INTRODUCTION

Generative adversarial networks (GANs) [1] have revolutionized AI by generating highly realistic synthetic data from limited datasets, addressing key challenges in data-scarce fields. They are widely used in computer vision for video and image synthesis, super-resolution transformations, and style transfer tasks, enabling photorealistic content generation and data augmentation. In medical imaging, GANs have been used to create synthetic scans to augment datasets, aiding diagnostic tools without the need for extensive real data [2]. GANs also play a crucial role in autonomous driving for generating training data and 3D object synthesis [3] for gaming and AR/VR applications. The adversarial training mechanism between a generator and discriminator allows GANs to closely mimic real data, significantly pushing the boundaries of AI applications where real data is limited or expensive to obtain.

Despite their impressive capabilities, the computational complexity and memory demands of GANs present significant challenges for traditional electronic accelerators. Unlike conventional convolutional neural networks (CNNs) [4], GANs introduce unique operations, such as transposed convolutions, which restructure the input matrix to upsample feature maps before applying a kernel, potentially leading to inefficient resource utilization on conventional architectures. This inefficiency is further exacerbated by layers like instance normalization (IN) [5], which require frequent memory accesses, adding to the strain on resource-constrained devices. As a result, accelerating GANs on traditional platforms often results in substantial energy and performance penalties.

To overcome these inefficiencies, silicon photonics has recently emerged as a groundbreaking solution to the limitations of conventional electronic architectures. As electronic accelerators face inherent limitations in the post-Moore era, including high fabrication costs and diminishing performance returns [6], the transmission of data over metallic wires presents significant bandwidth and energy bottlenecks. Silicon photonics, with its ultra-high bandwidth, low latency, and energy-efficient data communication, has emerged as a promising solution. In addition to replacing metallic wires for high-speed data transfers and supporting CMOS-compatible chip-level integration [7], silicon photonics has been shown to efficiently accelerate deep neural networks (DNNs) like CNNs, RNNs, LLMs, and GNNs [8]-[11]. By employing optical components, such as arrays of microring resonators (MRs) that function as matrix-vector multipliers in photonic integrated circuits (PICs), these accelerators enable low-latency, energy-efficient processing in the optical domain.

In this paper, we introduce *PhotoGAN,* the first silicon-photonic-based GAN accelerator that can accelerate the inference of a broad family of GAN models. Our novel contributions include:

- A reconfigurable GAN accelerator using non-coherent silicon photonics, capable of supporting transposed convolutions, instance normalization, and other GAN-specific layers, unlike existing photonic NN accelerators.
- Hardware and software co-design approaches, including sparse computation and scheduling optimizations, for efficient acceleration of diverse GAN models on the photonic-based hardware accelerator.
- A comprehensive comparison against GPU, CPU, TPU and state-of-the-art GAN accelerators, demonstrating superior performance and energy efficiency.

The remainder of this paper is organized as follows. Section II discusses the background and challenges associated with accelerating GANs. Section III details the *PhotoGAN* architecture and the innovations introduced to support efficient GAN inference. Section IV presents our experimental results. Lastly, Section V concludes with potential directions for future research.

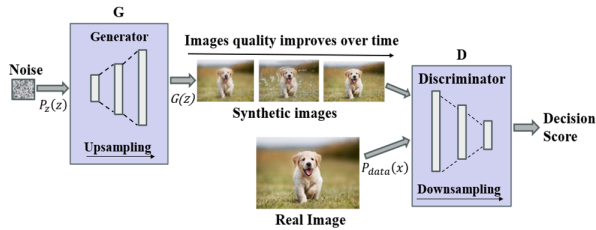

Fig. 1: High-level visualization of GAN model operation.

## II. BACKGROUND

### A. *Generative Adversarial Network (GAN) Models*

A GAN consists of two neural networks (see Fig. 1): a Generator (G) and a Discriminator (D), that engage in a competitive training process known as adversarial training [1]. This process can be viewed as a minimax game between two players, each trying to optimize opposing objectives. The generator aims to model the data distribution $P_{data}(x)$ and generate synthetic samples $G(z)$ where $z$ is a noise vector drawn from a known prior distribution $P_z(z)$ (typically Gaussian or uniform). The discriminator, in contrast, is tasked with distinguishing between real samples from the true data distribution and synthetic samples generated by the generator.

The generator's goal is to reduce the likelihood that the discriminator correctly identifies its outputs as fake, while the discriminator aims to improve its ability to distinguish real samples from fake ones. As shown in Fig. 1, the process begins with the noise vector $P_z(z)$ fed into the generator, which transforms it into a synthetic image. The discriminator then receives both the real $P_{data}(x)$ and generated images $G(z)$ to determine whether the synthetic one is real or fake. Over time, the generator becomes better at approximating the true data distribution, making its outputs increasingly realistic.

In this setup, the generator uses transposed convolutions to turn random noise into high-dimensional data, such as images, while the discriminator employs convolutional layers to downsample and classify inputs as real or synthetic. This adversarial training dynamic underpins the effectiveness of GANs in producing realistic outputs across various tasks. However, a key challenge is maintaining balance between the generator and discriminator. If one becomes too dominant, the other struggles to learn, potentially leading to issues like mode collapse, where the generator produces limited output variety.

### B. Hardware Acceleration of GANs

Prior efforts to accelerate GANs have primarily focused on optimizing specific layers or operations within the GAN framework. Although these approaches have demonstrated potential, they each possess limitations that impede their scalability and efficiency, particularly when addressing the computational and memory demands of modern GAN architectures. For example, a unified Multiple Instruction, Multiple Data (MIMD) - Single Instruction, Multiple Data (SIMD) architecture was proposed in [12] to accelerate transposed convolutions. The MIMD-SIMD approach aims to reorder computations and eliminate ineffectual operations associated with these transposed convolutions. However, managing both MIMD and SIMD execution increases design complexity and may not fully address memory bandwidth and latency issues, especially for large-scale GAN models.

An FPGA-based solution in [13] introduced an adaptable architecture capable of supporting various GAN model types. FPGAs provide flexibility and power efficiency, but they face limitations in terms of raw computational throughput compared to ASICs or specialized accelerators [14]. Additionally, the reconfiguration overhead and limited memory bandwidth of FPGAs restrict their utility in real-time, large-scale GAN inference tasks. ReRAM-based Process-in-Memory (PIM) architectures [15] offer another promising solution by reducing data movement between memory and processing units. ReRAM allows in-memory computation, alleviating memory bottlenecks that traditional accelerators face. However, ReRAM technology presents a distinct set of challenges such as reliability issues and significant area overhead [16], particularly when applied to large-scale GAN models with complex layers like transposed convolutions and instance normalization.

Other prior efforts have specifically targeted individual operations within GANs. For instance, [17] proposed an engine optimized for Instance Normalization (IN) layers, which are commonly used in style-transfer tasks. While this improves efficiency for tasks involving IN, it lacks generalizability across different GAN models that may use other types of normalization, or no normalization at all. Similarly, [18] focused on computation-efficient transformations, while [19] aimed to reduce off-chip memory access. Although these approaches improve dataflow and reduce memory traffic, they do not address broader computational challenges associated with large, multi-layered GAN models. Unlike previous approaches, *PhotoGAN* is the first GAN accelerator to leverage silicon photonics, offering a unified platform that enables high-throughput, low-latency data transfer and computation for GAN operations.

### C. Photonic Building Blocks for GAN Acceleration

Optical neural network accelerators can be classified into two main categories: coherent and non-coherent architectures. Coherent architectures use a single wavelength, where parameters are imprinted onto the optical signal's phase, enabling Multiply and Accumulate (MAC) operations through phase modulation. In contrast, non-coherent architectures leverage multiple wavelengths, imprinting parameters onto the amplitude of the optical signal. This enables parallel operations across different wavelengths, greatly enhancing throughput [10]. To the best of our knowledge, *PhotoGAN* is the first silicon photonic accelerator designed specifically for GAN models. Fig. 2 provides an overview of the fundamental devices and circuits required for computing with silicon photonics. The following are the main components needed:

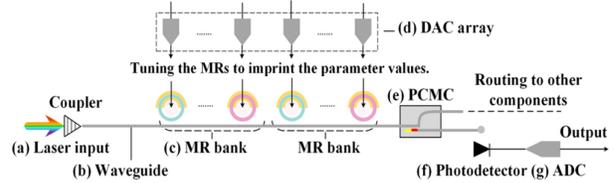

Fig. 2: Fundamental silicon photonic components.

*1) Lasers* generate the optical signals needed for computation and communication. Lasers can be on-chip, such as vertical cavity surface emission lasers (VCSELs), offering higher integration and lower losses, or off-chip, providing higher efficiency but suffering from coupling losses.

*2) Waveguides* carry the optical signals generated by the laser source. They are typically constructed with a high-refractive-index contrast material, such as a silicon (Si) core and a silicon dioxide (SiO$_2$) cladding, to enable total internal reflection. Using Wavelength Division Multiplexing (WDM), a single waveguide can support multiple wavelengths for parallel MAC operations (one operation per wavelength).

*3) Microring Resonators (MRs)* are optical modulators that can perform MAC operations by modulating input signals on their resonant wavelength. Each MR can be tuned to a specific wavelength ($\lambda_{MR}$), defined as $\lambda_{MR} = \frac{2\pi R}{m} n_{eff}$, where $R$ is the MR's radius, $m$ is the order of resonance, and $n_{eff}$ is the effective index of the device. MRs imprint input activations and weights for the matrix multiplication in GANs.

*4) Photodetectors (PDs)* convert the processed optical signals back into electrical signals. They must be sensitive enough to detect small input signals while compensating for optical losses (e.g., propagation, bending) along the link.

*5) Tuning circuits* control the effective index of MR devices through Thermo-Optic (TO) [20] or Electro-Optic

(EO) [21] tuning, enabling fine adjustments to the resonant wavelength coupling for error-free modulation.

*6) Digital-to-Analog Converters (DACs)* and *Analog-to-Digital Converters (ADCs)* are required for tuning the MRs and converting optical signals to the digital domain for intermediate processing. However, they are a major performance bottleneck in silicon photonic systems due to their high latency and energy costs.

*7) Phase Change Material Couplers (PCMCs)* leverage phase-change materials, which can switch between two states—amorphous and crystalline—with distinct optical properties. These materials enable non-volatile switching, meaning the state can be preserved without consuming power, thus significantly reducing static tuning power consumption. By applying a short optical or electrical pulse, the material can be switched between these states. This allows PCMCs to be used to dynamically route optical signals [7] between different blocks in photonic circuits, enabling reconfigurable and energy-efficient dataflows.

### D. Optical computations

The key operations in *PhotoGAN* are performed optically using the opto-electronic modulation devices, MRs. Fig. 3(a) illustrates the transmission plots for the input and through ports' wavelengths following the imprinting of a parameter onto the input signal. In many silicon-photonic systems, computations are executed by modifying the wavelength of a microring resonator ($\Delta\lambda_{MR}$), resulting in a predictable change in the amplitude of the optical signal's wavelength. *PhotoGAN* utilizes this principle to perform two primary computations with MR devices: summation and multiplication.

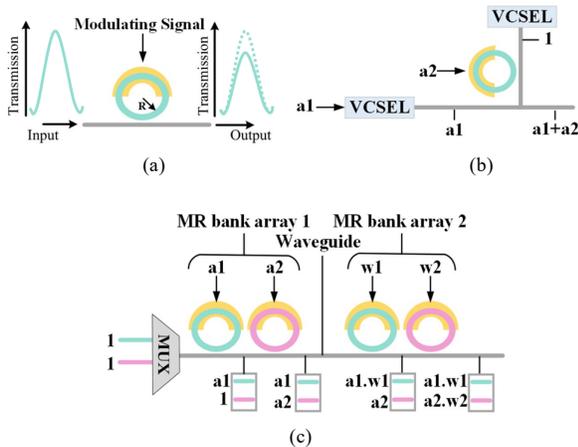

Fig. 3: (a) MR input and through ports' wavelengths after imprinting a parameter onto the signal; (b) MR device used to perform coherent summation of values a1 and a2; (c) MR bank arrays used to perform multiplication between input vector [a1, a2] and weight vector [w1, w2].

Summation is realized via coherent photonic summation, wherein optical signals with the same wavelength undergo constructive interference, resulting in the addition of their values. In Fig. 3(b), this is shown with two signals, $a1$ and $a2$. Using an analog biasing signal, VCSELs can be driven to produce an optical signal with a certain value imprinted onto it. Accordingly, the first value, $a1$, is imprinted onto the optical signal using the first VCSEL. The second VCSEL produces an optical signal of value 1, and $a2$ is imprinted onto the signal using an MR. When these two signals meet, their optical fields interfere constructively, yielding a new signal representing $a1 + a2$. Coherent summation is achieved through a laser phase-locking mechanism [22], which ensures that the VCSEL output signals are phase-aligned, enabling constructive interference. This summation is performed entirely in the optical domain, reducing the need for costly conversions between optical and electronic signals, thus boosting efficiency and reducing latency.

Multiplications, on the other hand, leverage multiple optical wavelengths that are directed through MR banks. Each MR bank imprints a specific activation or weight onto the optical signal. The result is a simultaneous multiplication of input activations and weights, performed in parallel across different wavelengths, and accumulated using a photodetector. In Fig. 3(c), we illustrate multiplication with two activations ($a1$ and $a2$) and two weights ($w1$ and $w2$) for simplicity. The first MR bank modulates the optical signals with activation values $a1$ and $a2$, while the second MR bank imprints the corresponding weight values $w1$ and $w2$ onto the same optical signals, resulting in a multiplication operation. These modulated signals are then passed through a photodetector (PD) to accumulate the result of the dot product ($a1w1 + a2w2$). This parallelization mimics the behavior of neurons in an artificial neural network (ANN), allowing matrix-vector multiplication (MVM) operations to be efficiently decomposed into vector multiplications.

## III. PHOTOGAN HARDWARE ACCELERATOR

*PhotoGAN* is a silicon photonic accelerator that can accelerate the inference of a broad family of GAN models. Fig. 4 shows an overview of the proposed architecture. The design consists of dense, convolutional, normalization, and activation blocks, with PCMCs utilized for the dynamic routing of optical signals between blocks. This approach significantly minimizes power consumption and latency associated with frequent opto-electronic data conversions. An integrated electronic control unit (ECU) performs tasks such as interfacing with the main memory, buffering intermediate results, and mapping matrices to the photonic domain. The following subsections describe the *PhotoGAN* architecture, and the hardware optimizations performed to efficiently accelerate GAN models. The proposed architecture consists of $L$ dense units and $M$ convolution units, designed to accelerate the various stages of GANs. Each dense and convolution block utilizes a single VCSEL array to supply the necessary optical signals across the rows in the MR bank arrays. This VCSEL reuse strategy not only minimizes the power consumption associated with laser sources but also reduces the potential for inter-channel crosstalk, ensuring the integrity of optical signals within the system.

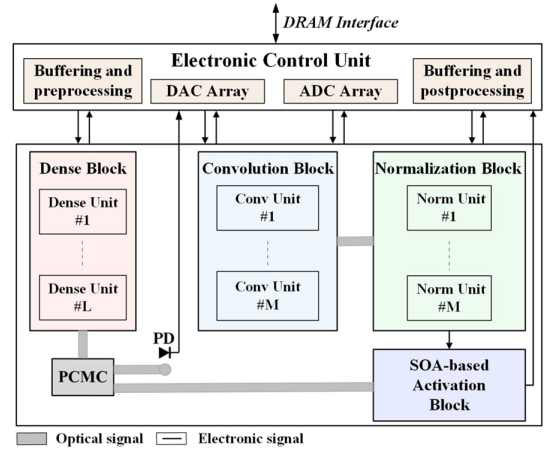

Fig. 4: Overview of *PhotoGAN* accelerator

## A. MR tuning circuit design

In *PhotoGAN*, precise tuning of MRs is essential for efficient optical computation. We assume a hybrid tuning circuit that combines EO [21] and TO tuning [20] methods to adjust the MR resonant wavelength ($\Delta\lambda_{MR}$). EO tuning, with low power consumption (≈4 µW/nm) and fast response times (≈ns range), is used for small wavelength adjustments. Alternatively, TO tuning offers a larger tunability range, but with the drawback of higher latency (≈µs range) and power consumption (≈27 mW/FSR). To further optimize power and reduce thermal crosstalk, we employ the Thermal Eigenmode Decomposition (TED) method [23]. TED ensures efficient TO tuning by minimizing interference between neighboring MRs, reducing overall power consumption. This hybrid tuning approach allows *PhotoGAN* to maintain high-speed and low-power operation across different GAN workloads.

## B. Architecture Design

### 1) Dense Block

The dense block comprises $L$ dense units, each optically implemented using two MR bank arrays with dimensions $K \times N$, where $K$ represents the number of rows and $N$ denotes the number of columns. These MR banks are responsible for executing MVMs utilizing the MR devices. As illustrated in Fig. 5, the first MR bank in each unit is responsible for the input activations, while the other imprints the weight values onto the optical signals. The modulated optical signals are then detected by Balanced Photodetectors (BPDs) at the end of each unit, producing a final accumulated analog value that represents the weighted sum of the inputs. BPDs are specialized PDs featuring two distinct arms connected to the same waveguide—one for positive signal polarities and the other for negative ones. This design enables them to handle both positive and negative parameter values by measuring the absolute difference between the two signals. The BPD independently sums the output signals from each arm, and subsequently, it calculates the net difference signal by subtracting the output of the negative arm from the positive arm. For adding the bias values, coherent summation is used where the MR banks' output drives a VCSEL unit operating at wavelength $\lambda_o$. Another VCSEL, also functioning at the same wavelength $\lambda_o$, is used to generate an optical signal with the bias values imprinted onto them. Accordingly, as explained earlier in Section II.D, as the two optical signals meet, the bias values will be added to the MVMs outputs. The final output is then routed directly to the subsequent activation block via a PCMC for further processing.

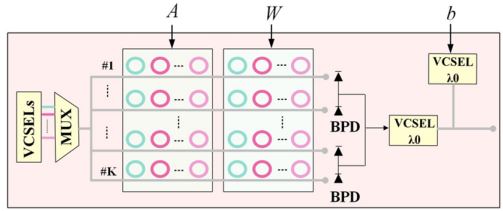

Fig. 5: Dense unit with two MR bank arrays, each of dimension $K \times N$.

### 2) Convolution Block

The convolution block comprises $M$ convolution units optimized for both convolution and transposed convolution operations. As discussed in [24], convolution operations can be transformed into vector multiplications to be accelerated using MR bank arrays. Each convolution unit contains two MR bank arrays, one for input activations and one for the weights, as illustrated in Fig. 6. These modulated optical signals are then transmitted directly to the normalization block for further processing, eliminating the need for costly intermediate opto-electronic conversions.

This architecture accelerates the convolutional operations by leveraging the parallelism of MR banks and the fast accumulation of signals via PDs, ensuring low-latency and energy-efficient computations.

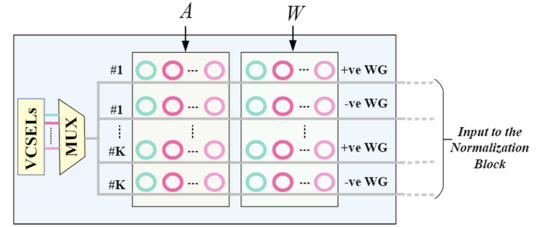

Fig. 6: Convolution unit consisting of two MR bank arrays, each with dimension $K \times N$.

### 3) Normalization Block

While Batch Normalization (BN) is commonly used in image generation tasks, generative models that focus on image-to-image translation, such as CycleGAN, typically favor instance normalization (IN) for improved performance. Unlike BN, where parameters are fixed after training, IN parameters are dynamically updated during both training and inference phases. *PhotoGAN* efficiently supports both types of normalization using broadband MRs [25]. The normalization block consists of $M$ normalization units, each designed to handle either BN or IN depending on the task requirements. The normalization parameters, which can be updated during inference, are used to tune the broadband MRs for real-time adjustment, as shown in Fig. 7. This flexible design allows *PhotoGAN* to adapt to different generative tasks while maintaining efficient optical processing using broadband MRs. Each normalization unit receives optical signals directly from the convolution units. In GANs, certain convolution layers require their outputs to pass through a normalization layer, while others do not. To accommodate this, as shown in Fig. 7, each normalization unit includes a mechanism to bypass broadband MRs, thereby avoiding the normalization operation when unnecessary.

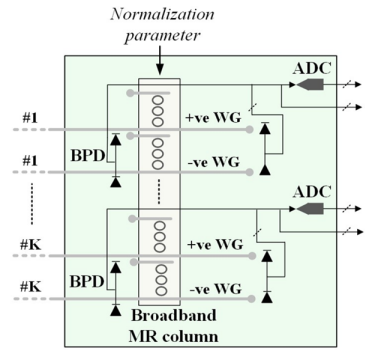

Fig. 7: Normalization unit consisting of Broadband MR column to imprint the normalization parameters.

### 4) Activation Block

The activation blocks comprise multiple activation units. Semiconductor-Optical-Amplifiers (SOAs) can be used to implement non-linear functions such as *Sigmoid* and *Tanh,* as demonstrated in [26]. Adjusting the gain of an SOA to a value close to one, will result in a RELU-like behavior. Expanding on this, we implement the Leaky ReLU activation function optically. The Leaky ReLU function can be illustrated as:

$$f(x) = \begin{cases} x & x > 0 \\ ax & x \leq 0 \end{cases} \quad (1)$$

As shown in Fig. 8, the Leaky ReLU block receives an input optical signal. A PD detects the analog value of the optical signal and passes it to a comparator circuit to determine the polarity of the signal. The output of the comparator circuit controls a PCMC switch. If the input signal to the Leaky ReLU block is positive, it is routed to an SOA tuned to 1 and if it is negative, the signal is fed into an SOA tuned to a small value "a". In general, by utilizing SOAs to implement the non-linear activation functions, the activation block efficiently processes both positive and negative inputs in the optical domain, reducing latency and power consumption compared to traditional electronic implementations.

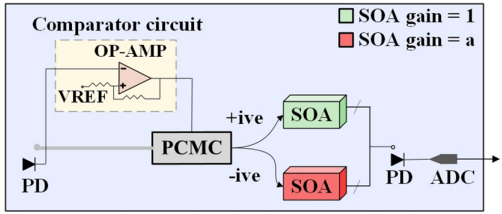

Fig. 8: SOA-based implementation of Leaky-ReLU.

*C. Dataflow and scheduling optimizations*

While optical-domain computations inherently deliver high throughput and energy efficiency gains, further optimization in both the hardware architecture design and the supporting software framework is essential. *PhotoGAN* incorporates three key GAN- and photonic-specific co-design optimizations to efficiently manage and schedule GAN operations: 1) sparse computation dataflow, 2) execution pipelining and scheduling, and 3) power gating. These optimizations are critical for improving memory bandwidth utilization, streamlining execution flow, and enabling a scalable, robust GAN acceleration solution.

*1) Sparse Computation Dataflow*

Efficient dataflow in GANs, particularly in the transposed convolution layers of the generator, plays a critical role in achieving high performance and energy efficiency in hardware accelerators. Fig. 9 illustrates the proposed optimization technique implemented in *PhotoGAN* for processing transposed convolution layers. In this example, a 3 × 3 filter with a stride of 1 and padding of 1 is applied to a 2 × 2 input feature map. Through zero-insertion, the input feature map is expanded to 5 × 5, and the filter is convolved over patches of the expanded input to generate the final output feature map, as shown in Fig. 9(a). Traditional convolution accelerators often encounter inefficiencies when handling transposed convolutions due to the insertion of zeros, which leads to the underutilization of computational resources. To overcome this, we propose a novel dataflow optimization technique specifically designed to meet the demands of transposed convolutions in GANs.

The technique works by first identifying all-zero columns in the flattened input feature map and the corresponding elements in the flattened kernel, as shown in Fig. 9(b). Once identified, these zero columns and their associated kernel elements are eliminated. This significantly reduces the complexity of the dot-product operation, as shown in Fig. 9(c) when computing the first output element. Accordingly, our proposed approach prevents the unnecessary injection of zeros into the *PhotoGAN* architecture, considerably reducing the amount of redundant computation and conserving valuable computational resources. To maintain data integrity and correct output dimensions, the removed columns are dynamically reintroduced during the computation stage in the ECU. This ensures that the final output remains accurate while optimizing both latency and power consumption.

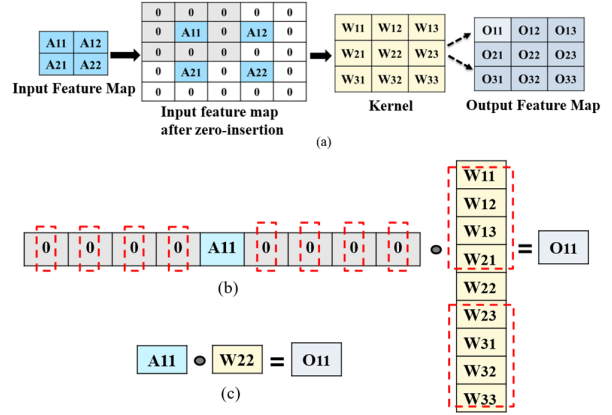

Fig. 9: (a) Illustration of transposed convolution operation. (b) Vector-dot product illustration for the first output element. The zeroes in the flattened input vector and corresponding kernel elements are highlighted. (c) Reduced dot-product operation.

*2) Execution Pipelining and Scheduling*

We employ pipelining at two levels of granularity. The first involves pipelining entire blocks of operations, where the dense block and its subsequent activation function are pipelined together, as depicted in Fig. 10(a). Similarly, the convolution block is pipelined with both the normalization and activation blocks, as shown in Fig. 10(b). The second pipelining technique in *PhotoGAN* is specific to the operations within the dense block, leveraging opto-electronic components. This block is organized into two distinct stages: the first stage incorporates DACs, VCSELs, and MR banks for the MVM operations, while the second stage employs PDs and VCSELs to the add bias values (see Fig. 5).

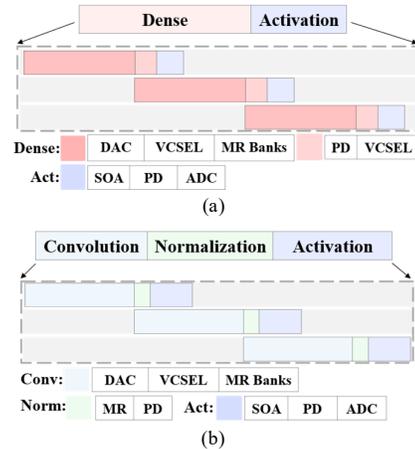

Fig. 10: (a) Pipelining of dense and activation layers. (b) Pipelining of convolution, normalization, and activation layers

The pipelining approach ensures that once the optical signals, with the MVM outputs encoded onto them are detected by the PDs, the next iteration of operations in the first stage can immediately begin. By executing these stages concurrently and in a streamlined manner, *PhotoGAN* achieves further reductions in latency while maintaining efficient overall power usage, optimizing performance without compromising energy efficiency.

*3) Power Gating*

We enable support for power gating as a key optimization strategy to enhance both latency and power efficiency. This technique ensures that only the active processing block is powered at any given time—when the dense block is active, the convolution block is deactivated, and vice versa. By selectively powering components as needed, *PhotoGAN* minimizes energy waste in idle parts. Additionally, this enables the sharing of DAC arrays between the dense and convolution blocks. The DAC devices are known for their high latency and energy demands in silicon photonic systems. Thus, this sharing approach not only reduces power consumption and latency but also eliminates unnecessary hardware duplication, leading to a more compact and energy-efficient design.

## IV. EXPERIMENTS AND RESULTS

We performed detailed analysis of the proposed *PhotoGAN* architecture, using four GAN models as specified in Table 1. To evaluate them, we employed the Inception Score (IS) metric [27], which assesses the quality of the images generated by the GAN. Our analysis indicated that 8-bit model quantization results in minimal degradation of IS, compared to full precision models. We thus targeted acceleration of 8-bit precision GAN models. The datasets used to evaluate the models and the percentage change in IS after 8-bit quantization is reported in Table 1. Tensorflow 2.9 was used to train and evaluate the models. For estimating the performance and energy costs of accelerating each GAN model using our proposed accelerated design, we developed a comprehensive simulator with optoelectronic device models aggregated to create a simulatable architectural model.

TABLE 1: EVAULATED MODELS, PARAMETERS AND IS

| Model | Dataset | Parameters | % change in IS after 8-bit quantization |
|---|---|---|---|
| DCGAN [28] | celebA | 3.98M | + 0.11 % |
| Cond. GAN [29] | F-MNIST | 1.17M | + 0.10 % |
| ArtGAN [30] | Art Portraits | 1.27M | - 6.64 % |
| CycleGAN [31] | Horse2zebra | 11.38M | - 0.36 % |

The optoelectronic devices' latencies and power characteristics considered in this work are shown in Table 2. Factors contributing to photonic signal losses, such as waveguide propagation loss ($1\ dB/cm$), splitter loss ($0.13\ dB$ [32]), combiner loss ($0.9\ dB$ [32]), MR through loss ($0.02\ dB$ [33]), MR modulation loss ($0.72\ dB$ [34]), EO tuning loss ($6\ dB/cm$ [21]), and TO tuning power ($27.5\ mW/FSR$ [20]) are taken into account.

TABLE 2: OPTOELECTRONIC PARAMETERS

| Devices | Latency | Power |
|---|---|---|
| EO Tuning [21] | 20 ns | 4 µW |
| TO Tuning [20] | 4 µs | 27.5 mW/FSR |
| VCSEL [9] | 0.07 ns | 1.3 mW |
| Photodetector [9] | 5.8 ps | 2.8 mW |
| SOA [9] | 0.3 ns | 2.2 mW |
| DAC (8-bit) [35] | 0.29 ns | 3 mW |
| ADC (8-bit) [36] | 0.82 ns | 3.1 mW |

The laser power required for each source in the architecture was modeled as:

$$P_{laser} - S_{detector} \geq P_{photoloss} + 10 \times \log_{10} N_\lambda \quad (2)$$

where $P_{laser}$ is the laser power in $dBm$, $S_{detector}$ is the photodetector sensitivity in $dBm$, $N_\lambda$ is the number of laser sources/wavelengths, and $P_{photoloss}$ is the total optical loss encountered by the signal. Minimizing errors during optical operations is essential to ensure reliable performance. Our photonic device-level analysis based on FDTD, CHARGE, MODE solver, and INTERCONNECT tools [38] shows that a waveguide can accommodate up to 36 MRs for non-coherent operation, while still maintaining error-free performance and minimizing crosstalk between optical wavelengths. *PhotoGAN* adheres to this guideline by keeping the number of MRs per waveguide below 36 in each operational unit.

In the next subsection, an analysis to determine the exact optimal values for *PhotoGAN's* architectural parameters discussed in Section III, including, *N, K, L* and *M*, is presented.

*A. PhotoGAN architectural design space exploration*

The performance of *PhotoGAN* is highly dependent on the values of four key parameters: *N* (the number of columns in each MR bank array), *K* (the number of rows in each MR bank array), *L* (the number of units in the dense block), and *M* (the number of units in the convolution and normalization blocks). These parameters significantly affect the overall throughput, latency, and energy efficiency.

We performed a comprehensive architectural design space exploration to identify the optimal configuration that maximizes GOPS/EPB (Giga-operations-per-second per Energy-per-bit). For this analysis, we set a power limit of $100\ W$, which aligns with our goal of creating an energy-efficient GAN accelerator suitable for resource-constrained environments. As shown in Fig. 11, the exploration yielded the most optimal configuration for [*N, K, L, M*] as [*16, 2, 11, 3*]. This is the optimal configuration as it offers the best balance between computational performance and power efficiency, by achieving the maximum GOPS/EPB value among all the points considered in the design space. This results in significant increase in throughput while keeping the power consumption within acceptable limits.

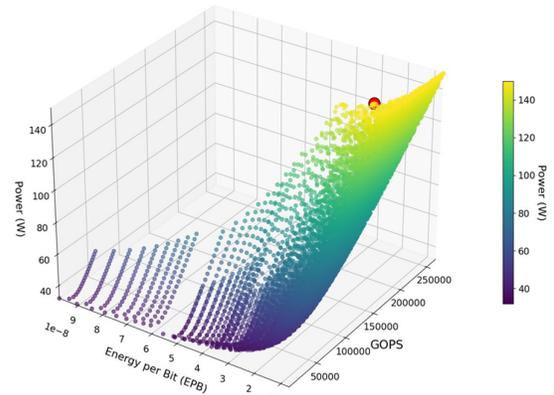

Fig. 11: Design space exploration for *PhotoGAN* architecture. The most optimal point is shown with a red dot.

*B. Orchestration and scheduling optimization analysis*

We conducted a sensitivity analysis to assess the impact of each of the dataflow and scheduling optimizations described in Section III. The normalized energy results are shown in Fig. 12. The sparse computation dataflow (referred to as S/W Optimized), Pipelining, Power Gating and the proposed approach with all the optimizations combined (S/W Optimized + Pipelined + Power Gating) were explored in this analysis. The Baseline configuration, which does not incorporate any of the optimizations discussed, serves as the reference point for normalizing the results. As shown in Fig. 12, each optimization technique significantly reduces energy consumption compared to the baseline. On average, the combined optimizations of S/W Optimized, Pipelining, and Power Gating result in a 45.59× reduction in normalized energy consumption across all models vs. the baseline.

Notably, the impact of each optimization varies across the different GAN models, offering valuable insights into their distinct effects. For instance, CycleGAN consists of fewer transposed convolution layers compared to the other GAN models. This explains why the S/W Optimized approach, with the sparse computation dataflow enabled, has a less pronounced effect on energy reduction in CycleGAN compared to the other models. In contrast, the Pipelined optimization, which efficiently managed the dataflow between the CycleGAN's various other layers, shows a more substantial energy reduction. Overall, the results demonstrate the effectiveness of integrating software and hardware optimizations to significantly reduce the energy consumption of GAN inference in our *PhotoGAN* architecture, enhancing the feasibility of deploying these models in energy-constrained environments.

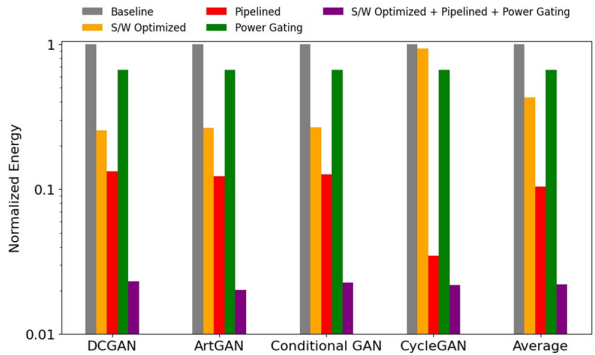
Fig. 12: Energy improvements due to dataflow and scheduling optimization.

### C. Comparison to state-of-the-art accelerators

We compared *PhotoGAN* with the baseline A100 GPU, Intel Xeon CPU, TPU v2 [37], FPGA-based GAN accelerator from [13] and ReRAM-based GAN accelerator from [15]. We used our simulator to estimate the Giga Operations Per Second (GOPS) and Energy Per Bit (EPB) for each model. Fig. 13 shows the GOPS comparison between the different platforms. *PhotoGAN* achieved on average 134.64×, 260.13×, 123.43×, 286.38× and 4.40× higher GOPS than GPU, CPU, TPU, FPGA-based accelerator and ReRAM-based accelerator, respectively.

The high GOPS performance of *PhotoGAN* is primarily due to its use of optical computations through silicon photonics, which offer high bandwidth and low latency, enabling efficient parallelism and rapid data processing. These advantages are further enhanced by the hardware and software optimizations that align data transfer and computation, minimizing idle time and maximizing throughput. In doing so, *PhotoGAN* effectively overcomes the limitations of traditional CPU and GPU architectures, which struggle with memory stalls and inefficient data movement; FPGA and TPU accelerators, which encounter bandwidth bottlenecks due to electronic interconnects; and ReRAM-based accelerators, which, despite their energy efficiency, are hindered by slow non-volatile memory access times. As a result, *PhotoGAN* achieves significant performance improvements over prior efforts.

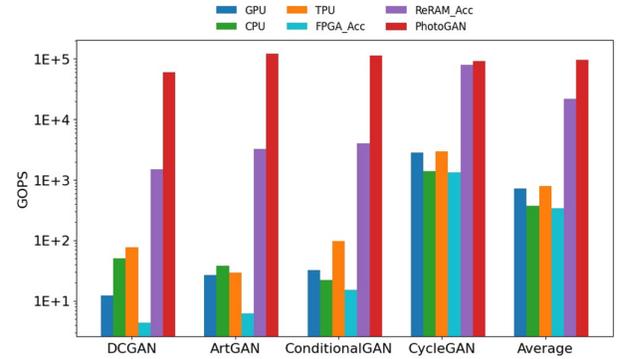
Fig. 13: GOPS comparison across different GAN models.

Fig. 14 shows the EPB comparison between *PhotoGAN* and the other architectures considered. *PhotoGAN* achieves on average 514.67×, 60×, 313.50×, 317.85× and 2.18× lower EPB when compared to GPU, CPU, TPU, FPGA-based accelerator and ReRAM-based accelerator, respectively. These EPB improvements can be attributed to *PhotoGAN*'s low-latency operations and energy-efficient computational flow. Moreover, it is important to note that *PhotoGAN* employs flexible energy-saving techniques that can be customized for the specific model being accelerated. These techniques include power gating of unused blocks and the reuse of VCSEL and DAC arrays in the photonic domain, resulting in a significant reduction in overall power consumption. In contrast, traditional electronic platforms like GPUs and CPUs struggle to implement effective power gating due to the overhead of managing multiple general-purpose cores. Similarly, ReRAM accelerators are limited by slower switching speeds and higher write energies for memory access, whereas photonic systems, with optical data transfers, minimize these overheads.

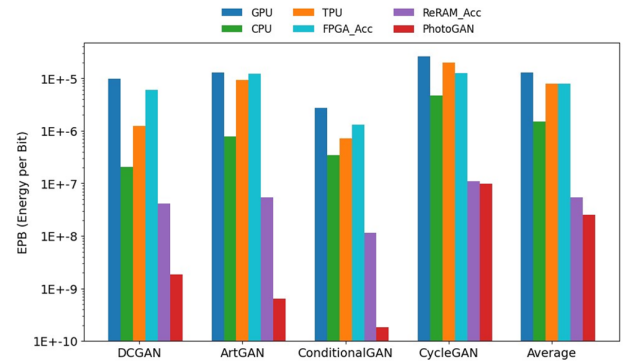
Fig. 14: EPB comparison across different GAN models.

### V. CONCLUSION

In this paper, we presented *PhotoGAN*, a novel silicon photonic-based accelerator that targets increasingly ubiquitous GAN models. In comparison to five computing platforms and state-of-the-art GAN accelerators, *PhotoGAN* exhibited throughput improvements of at least 4.4× and energy-efficiency improvements of at least 2.18×. These results showcase the potential of *PhotoGAN* to offer high-throughput, energy-efficient inference acceleration for GAN models. While our focus was on optimizing the hardware architecture, future work can explore the integration of advanced software optimization techniques to further enhance the throughput and energy efficiency. Additionally, addressing emerging challenges with silicon photonics, such as fabrication process variations [39], integration with non-volatile memory solutions [40], and security vulnerabilities [41] could lead to further improvements in performance and scalability, towards practical deployments.